\documentclass[iop]{emulateapj}

\shorttitle{Formation of Super-Earths by Giant Impacts}
\shortauthors{Matsumoto \& Kokubo}

\begin{document}

\title{Formation of Close-in Super-Earths by Giant Impacts: Effects of Initial Eccentricities and Inclinations of Protoplanets}

\author{Yuji Matsumoto,\altaffilmark{1,2} \& Eiichiro Kokubo\altaffilmark{2,3}}
\email{ymatsumoto@cfca.nao.ac.jp}

\altaffiltext{1}{Planetary Exploration Research Center, Chiba Institute of Technology, Narashino, Chiba, 275-0016, Japan}
\altaffiltext{2}{Center for Computational Astrophysics, National Astronomical Observatory of Japan, Osawa, Mitaka, Tokyo, 181-8588, Japan}
\altaffiltext{3}{Division of Theoretical Astronomy, National Astronomical Observatory of Japan, Osawa, Mitaka, Tokyo, 181-8588, Japan}

\begin{abstract}

Recent observations have revealed the eccentricity and inclination distributions of close-in super-Earths.
These distributions have the potential to constrain their formation processes.
In the in-situ formation scenario, the eccentricities and inclinations of planets are determined by gravitational scattering
 and collisions between protoplanets on the giant impact stage.
We investigate the effect of the initial eccentricities and inclinations of protoplanets on the formation of close-in super-Earths.
We perform $N$-body simulations of protoplanets in gas-free disks, changing the initial eccentricities and inclinations systematically. 
We find that while the eccentricities of protoplanets are well relaxed through their evolution, the inclinations are not.
When the initial inclinations are small, they are not generally pumped up since scattering is less effective
 and collisions occur immediately after orbital crossing.
On the other hand, when the initial inclinations are large, they tend to be kept large since collisional damping is less effective.
Not only the resultant inclinations of planets, but also their number, eccentricities, angular momentum deficit,
 and orbital separations are affected by the initial inclinations of protoplanets.

\end{abstract}


\section{INTRODUCTION}\label{sect:intro}

In recent years, radial velocity (RV) and transit observations have discovered a large number of super-Earths
 \citep[$\lesssim 30M_{\oplus}$ or $\leq 6R_{\oplus}$, where $M_{\oplus}$ and $R_{\oplus}$ are the mass and radius of the Earth,
 respectively, e.g.,][]{Mayor+2011, Mullalley+2015}. 
Most of the observed super-Earths are distributed within $\simeq0.3$ au. 
While some excesses are seen around first-order commensurabilities in the period distribution,
 super-Earths are normally not in mean-motion resonances \citep[e.g.,][]{Lissauer+2011, Baruteau&Papaloizou2013}. 
Super-Earths observed by the RV method have eccentricities, $e\lesssim0.4$ \citep[e.g.,][]{Mayor+2011},
 while those observed by the {\it Kepler} mission have small eccentricities, $e\sim 0.01-0.1$
 \citep{Fabrycky+14,Hadden&Lithwick2014, VEylen&Albrecht2015}. 
When a planetary system has more than one transiting planet, the mutual inclination $i$ can be estimated.
The mode of the mutual $i$ of transiting systems discovered by {\it Kepler} is 0.017 rad $<i<$ 0.039 rad \citep{Fabrycky+14}. 

The accretion process of close-in super-Earths has been studied using $N$-body simulations and analytical calculations.
\cite{Hansen&Murray12,Hansen&Murray13} investigated the accretion of close-in super-Earths from protoplanets or planetary embryos.
\cite{Tremaine2015} calculated analytically the accretion process.
\cite{Dawson+2016} considered the accretion from protoplanets including the effect of gas drag.
\cite{Moriarty&Ballard2016} performed the simulations from the initial disk consisting of protoplanets and planetesimals.
\cite{Ogihara+2015, Ogihara+2015b} considered the interactions of  protoplanets with nebula gas and planetesimals.
Especially, the typical eccentricity and its distribution are compared to the observation in \cite{Hansen&Murray13},
 \cite{Tremaine2015} and \cite{Ogihara+2015, Ogihara+2015b}.
In \cite{Hansen&Murray13} and \cite{Tremaine2015}, the typical $e$ is about $0.1\pm0.05$, which is comparable to the observed value.
\cite{Ogihara+2015, Ogihara+2015b} found that the distributions of the observed period ratios and eccentricities are reproduced
 under the nebula gas evolution with the disk wind.
However, the observed mass distribution is not reproduced by their simulations.
It still remains important to know how eccentricities and inclinations of planets are dynamically determined.

On the giant impact stage, the eccentricities of protoplanets evolve through gravitational scattering and collisions
 between protoplanets in gas-free environments.
The eccentricity evolution of protoplanets by collisions was studied by \cite{Matsumoto+2015}. 
Close-in protoplanets collide soon after their orbits cross, where their eccentricities are about the value needed for orbit crossing
 and their pericenters are located in the opposite direction. 
By close approach, the osculating eccentricities increase to about the escape eccentricities,
 defined by $e_{\rm esc}=v_{\rm esc}/v_{\rm K}$, where $v_{\rm esc}$ and $v_{\rm K}$ are the escape and Kepler velocities, respectively. 
These eccentricities are damped by the collision since the random velocities, which are the velocities relative to the circular orbit on the midplane
 given as $\simeq \sqrt{e^2+i^2}v_{\rm K}$ \citep[e.g.,][]{Lissauer&Stewart1993}, are canceled due to the opposite direction of their pericenters. 
The final planets have smaller random velocities than the surface escape velocities. 
They also indicated that the eccentricity-damping rate depends on the initial eccentricities of protoplanets, 
which suggests that the final random velocities of planets are also affected by the initial random velocities of protoplanets.

\cite{Kokubo&Ida07} and \cite{Dawson+2016} investigated the effects of the initial $e$ and $i$ of protoplanets
 on the basic properties of planets formed by giant impacts. 
\cite{Kokubo&Ida07} studied accretion of protoplanets around 1 au. 
In their simulations, the initial $e$ and $i$ are given by the Rayleigh distribution with dispersions $\langle e^2\rangle^{1/2} = 2\langle i^2\rangle^{1/2} = 0.01$
 and $\langle e^2\rangle^{1/2} = \langle i^2\rangle^{1/2} = 0.01$ (the unit of $i$ is radian). 
Within this range of the initial values, they found that the initial $e$ and $i$ of protoplanets barely affect the basic properties of planets 
 since the initial $e$ and $i$ are well relaxed during the giant impact stage. 
However, \cite{Kokubo&Ida07} explored only the limited range of the initial $e$ and $i$ around 1 au. 
\cite{Dawson+2016} also studied accretion of protoplanets. 
The innermost protoplanets in their simulations are located around 0.05 au. 
They changed the initial orbital width of protoplanets $ea$ from 0 to $0.5\Delta a/a$, where $\Delta a$ is the orbital separation of adjacent protoplanets. 
They concluded that the initial eccentricities affect the final orbital separations. 
On the other hand, they also pointed out that their results for the $0.5\Delta a/a$ case are different from those for the other initial eccentricity cases. 
They also varied the initial inclinations, given from a uniform distribution between $0-0.001^{\circ}$ and $0-0.1^{\circ}$. 
The median inclination grows to equipartition with eccentricity within a fraction of a Myr, 
and modest differences in the final inclinations and separations exist. 
They explained the relation between the initial inclinations and the final separations by the change of stirring and merger rates. 

These studies showed the effects of the initial $e$ and $i$ of protoplanets on the final assemblage of planets.
However, the elementary process of orbital evolution due to gravitational scattering and collisions is still unclear.
In particular the inclination-damping rate by a collision is unknown. 
\cite{Matsumoto+2015} showed that the inclination-damping rate is different from the eccentricity one.
This is because collisions occur when eccentricities become larger than $0.5\Delta a/a$, independently of inclinations. 

In this paper, we study the effects of the initial $e$ and $i$ of protoplanets on the final orbits of planets formed by giant impacts. 
Especially, we focus on the collisional evolution of $e$ and $i$.
In Section \ref{sect:model}, we describe our numerical model and method. 
In Section \ref{sect:results}, we present results.
Section \ref{sect:summary} is devoted to a summary and discussion.

\section{NUMERICAL MODEL}\label{sect:model}

\subsection{Initial Conditions}

We perform $N$-body simulations of the protoplanet accretion without small planetesimals 
or any damping forces to investigate the basic dynamics of the giant impact stage. 
We perform three sets of simulations with different initial numbers of protoplanets in which the total mass ($M_{\rm tot}$)
 and initial mean semimajor axis ($\langle a\rangle$) are approximately the same, namely, 17$M_{\oplus}$ and 0.13 au, respectively.
The numbers of initial protoplanets ($N$) are 16, 8, and 32. 
The protoplanets have the isolation masses ($M$) \citep{Kokubo&Ida02}, orbiting around a solar mass ($M_{\odot}$) star.
The surface density of protoplanets is given by $\Sigma = \Sigma_1 (a/{\rm 1\ au})^{-3/2}$, where $\Sigma_1$ is the surface density at 1 au.
The orbital separations are fixed as $\Delta a = 9r_{\rm H}$ - $10 r_{\rm H}$, where $r_{\rm H}$ is the mutual Hill radius
 given by $r_{\rm H} = [(M_j+M_{j+1})/3M_{\odot}]^{1/3} (a_j+a_{j+1})/2$, where $j$ means the $j$-th protoplanet from the innermost.
Then, the semimajor axis of the innermost protoplanet ($a_1$) is different in the three models, and the systems are compact when $N$ is small.
In N16 models, the innermost protoplanet is located at 0.05 au.
The bulk density of protoplanets is given as $\rho=3\ {\rm g\ cm^{-3}}$.
The initial $e$ and $i$ of protoplanets are given by the Rayleigh distribution. 
The inclinations are defined from the system invariable plane.
Their ranges are $\langle e_{\rm ini}^2\rangle^{1/2} = 10^{-3}$ - $10^{-1}$ and $\langle i_{\rm ini}^2\rangle^{1/2} = 10^{-3.6}$ - $10^{-0.5}$ radian.
Note that $\langle e^2 \rangle^{1/2}$ and $\langle i^2 \rangle^{1/2}$ are changed independently,
 not following $\langle e^2 \rangle^{1/2} = 2 \langle i^2 \rangle^{1/2}$. 
In our models, $N$, $\langle e_{\rm ini}^2\rangle^{1/2}$, and $\langle i_{\rm ini}^2\rangle^{1/2}$ are the independent variables.
The models are named by these parameters, such as N16e-1.5i-1, where initially 16 protoplanets have $\langle e_{\rm ini}^2\rangle^{1/2}=10^{-1.5}$
 and $\langle i_{\rm ini}^2\rangle^{1/2}=10^{-1}$ rad.
The initial conditions of the protoplanet systems are summarized in Table \ref{table:initial}. 

\subsection{Orbital Integration}

The orbits of protoplanets are calculated numerically by integrating the equation of motion of protoplanets,
\begin{eqnarray}
	\frac{{\rm d}^2 \textrm{\boldmath $r$}_i}{{\rm d}t^2}
		= -{\rm G}M_{\odot}\frac{\textrm{\boldmath $r$}_i}{|\textrm{\boldmath $r$}_i|^3}
		- \sum_{ j \neq i} {\rm G} M_{j} \frac{ \textrm{\boldmath $r$}_i - \textrm{\boldmath $r$}_j }{|\textrm{\boldmath $r$}_i - \textrm{\boldmath $r$}_j |^3}
		- \sum_j {\rm G} M_{j} \frac{\textrm{\boldmath $r$}_j}{|\textrm{\boldmath $r$}_j|^3},
\end{eqnarray}
where $\textrm{\boldmath $r$}$ is the position of protoplanets.
We adopt the fourth-order Hermite integrator \citep[e.g.,][]{Kokubo&Makino2004} with the hierarchical timestep \citep{Makino1991}. 
The simulations follow the evolution of protoplanet systems for $\gtrsim 10^8 t_{\rm K}$ of the innermost protoplanet,
 where $t_{\rm K}$ is the Kepler time and $10^8 t_{\rm K}\sim10^6$ yr when $a_1=0.05$ au. 
We estimate the orbital crossing timescale of final planetary systems using the empirical equations \citep[e.g.,][]{Chambers+1996, Yoshinaga+1999, Zhou+2007}.
The crossing timescale is the time until a close encounter between planets with distance $< r_{\rm H}$, occurs.
We confirm that the crossing timescales of final systems are longer than $10^8 t_{\rm K}$ by the estimation in \cite{Ida&Lin2010} or numerical integration.
We calculate 20 runs for each model of protoplanet systems, changing initial orbital angles of protoplanets randomly. 

The perfect accretion is assumed for collisions.
This assumption holds when the collisional velocity is smaller than the escape velocity of protoplanets
 \citep[e.g.,][]{Genda+2012, Stewart&Leinhardt2012, Genda+2015}. 
Since the protoplanets in close-in orbits have a relatively short collision timescale,
 it is expected that protoplanets collide soon after their orbits cross.
In such collisions, the collision velocities are $\simeq v_{\rm esc}$ and collisions usually result in accretion. 
We discuss the validity of the assumption of the perfect accretion in section \ref{sect:results}.

\section{RESULTS}\label{sect:results}

\subsection{Effect of Initial Eccentricities}\label{sect:e}

\label{sect:e_init}

\begin{figure}[htpb]
	\epsscale{1}
	\plotone{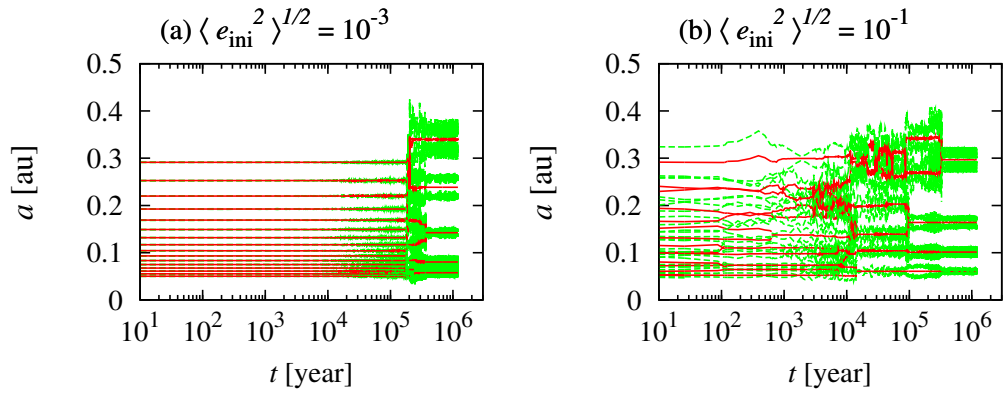}
	\caption{
		Typical orbital evolution of protoplanets in models N16e-3i-3.6 (a) and N16e-1i-3.6 (b).
		The time evolution of the semimajor axes (red solid lines), and pericenter and apocenter distances (green dashed lines) is plotted.
	}
	\label{fig:ap_runA_e}
\end{figure}

First we investigate the effect of the initial eccentricities of protoplanets on the collisional evolution of orbits
 and the final structure of planetary systems.
We show an example of the typical evolution of the small eccentricity model (model N16e-3i-3.6) in panel (a) of Figure \ref{fig:ap_runA_e}.
In this model, protoplanets initially have $\langle e_{\rm ini}^2 \rangle^{1/2} = 10^{-3}$ and
 $\langle i_{\rm ini}^2 \rangle^{1/2} = 10^{-3.6}$ rad.
These eccentricities correspond to $\lesssim0.1h$, where $h$ is the reduced Hill radius of protoplanets,
 $h = r_{\rm H}/a $.
Since $\langle e_{\rm ini}^2 \rangle^{1/2}$ is very small, the orbital crossing timescale is longer
 and collisions do not occur in $\lesssim10^5$ year \citep{Yoshinaga+1999, Zhou+2007, Pu&Wu2015}. 
After the first collision occurs at $1.8\times 10^5$ year, collisions occur one after another,
 and $N$ of protoplanets decreases rapidly. 
In this run, the final collision is at $3.7\times10^5$ year. 
Finally, five planets are formed.
Their orbital separations are $\geq 20r_{\rm H}$ and the orbital crossing time calculated by the estimation in \cite{Ida&Lin2010} is longer than $5\times10^{9}t_{\rm K}$. 

\begin{figure}[htpb]
	\plotone{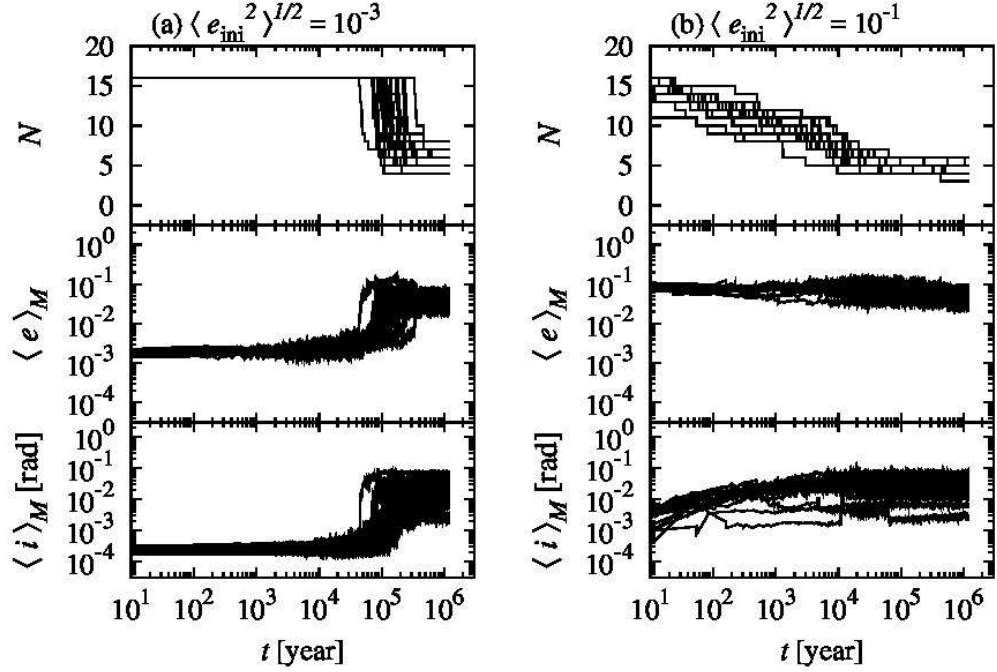}
	\caption{
	Time evolution of the number of protoplanets ($N$), mass-weighted eccentricity ($\langle e \rangle_M$), and mass-weighted inclination
	 ($\langle i \rangle_M$) in models N16e-3i-3.6 (a) and N16e-1i-3.6 (b) are plotted for all runs.
}
\label{fig:Time_ev_N+massweighted_e_i_runA_e}
\end{figure}

The time evolution of the number and mass-weighted mean eccentricity and inclination ($\langle e \rangle_M,\ \langle i \rangle_M$)
 of protoplanet systems in all the runs of model N16e-3i-3.6 are shown in panel (a) of Figure \ref{fig:Time_ev_N+massweighted_e_i_runA_e}. 
The first collisions occur within $\sim 10^5\ {\rm year}$ and $N$ decreases rapidly, while $\langle e \rangle_M$ and $\langle i \rangle_M$
 significantly increase around $\sim10^5$ years.
This is because $e$ and $i$ of protoplanets increase by close scattering.
After protoplanets repeat collisions in $\sim10^5$ yr, the final planets have $10^{-2}<\langle e \rangle_M<10^{-1}$,
 which is smaller than $e_{\rm esc}$ due to collisional damping \citep{Matsumoto+2015}. 
The final $\langle i \rangle_M$ is between $10^{-3}$ rad and $10^{-1}$ rad and its dispersion is larger than that of $\langle e \rangle_M$
 (see \S\ref{sect:col_damp_e}, \ref{sect:col_damp_i}). 

Panels (b) of Figures \ref{fig:ap_runA_e} and \ref{fig:Time_ev_N+massweighted_e_i_runA_e} also shows the results of the large eccentricity model (model N16e-1i-3.6).
The initial eccentricities correspond to $5.7 h - 8.6 h$ of the initial protoplanets. 
Since the initial orbital separations are $\sim9r_{\rm H}$, the radial excursions of the protoplanets ($ea$) are comparable to the orbital separations. 
In such a large eccentricity case, orbital crossing occurs immediately.
The first collisions occur within $\lesssim 10\ {\rm year}$, and $N$ decreases almost linearly with logarithmic time ($\log{t}$) until around $\sim10^4$ year.
Whereas the evolution timescale is different between the two models (models N16e-1i-3.6 and N16e-3i-3.6), the final planetary systems are not so different.
This means that most of the initial difference of $e$ is relaxed through scattering and collisions \citep{Matsumoto+2015}. 

\subsubsection{Orbital Evolution}\label{sect:col_damp_e}

Next we focus on the evolution of orbital elements at collisions.
\cite{Matsumoto+2015} showed that the osculating $e$ of colliding protoplanets just before a collision is $e \simeq e_{\rm esc}$,
 which is an outcome of their close approach.
This eccentricity is damped by the collision. 
Since orbital crossing initially occurs at around the apocenter of the inner body and the pericenter of the outer body,
 the difference between the pericenters in a collision ($\Delta \varpi$) is about 180$^{\circ}$. 
The velocity of the planet at pericenter is larger than the local circular velocity and that at apocenter is smaller. 
Then, the velocity of the merged body becomes closer to the circular velocity than those of the colliding bodies,
 i.e., the merged body has smaller $e$ than those of the colliding bodies. 

\begin{figure}[htpb]
	\plotone{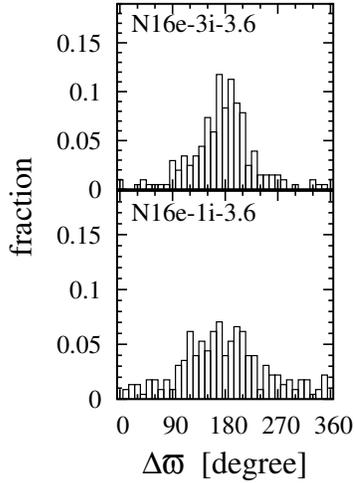}
	\caption{
		Distribution of the difference of the longitude of the pericenter ($\Delta \varpi$) of colliding bodies just before the collision in models N16e-3i-3.6
		 (top) and N16e-1i-3.6 (bottom).
		The total numbers of collisions are 204 in model N16e-3i-3.6 and 227 in model N16e-1i-3.6.
		The mean and variance of $\Delta \varpi$ are $179^{\circ}\pm 54.5^{\circ}$ in model N16e-3i-3.6 and $178^{\circ}\pm 74.3^{\circ}$ in model N16e-1i-3.6.
	}
\label{fig:sort_dvarpi_e}
\end{figure}

Figure \ref{fig:sort_dvarpi_e} shows the $\Delta \varpi$ distribution of two colliding bodies just before a collision in models N16e-1i-3.6 and N16e-3i-3.6. 
In both models the $\Delta \varpi$ distribution is peaked around $180^{\circ}$. 
The variance of $\Delta \varpi$ in model N16e-1i-3.6 is $74.3^{\circ}$ and that in model N16e-3i-3.6 is $54.5^{\circ}$. 
Since $\langle e_{\rm ini}^2 \rangle^{1/2}$ in model N16e-1i-3.6 is larger than that in model N16e-3i-3.6,
 the early collisions in model N16e-1i-3.6 are less concentrated around $180^{\circ}$ than those in model N16e-3i-3.6. 
This difference is relaxed after several collisions and the $\Delta \varpi$ distributions become similar. 

On the other hand, the osculating $i$ just before a collision does not always become as large as $i_{\rm esc}=0.5e_\mathrm{esc}$.
When $i\lesssim h$, gravitational scattering takes place in the disk plane and $i$ increases slowly \citep{Ida1990}.
Since $\langle i_{\rm ini}^2\rangle^{1/2}$ is much smaller than $h\sim 0.01$ in models N16e-1i-3.6 and N16e-3i-3.6,
 $i$ does not reach $i_{\rm esc}$ by close scattering. 
While $e$ becomes larger than $h$ just before collisions, $i$ can keep small values.
Although $i$ is not always pumped up by close approach of colliding bodies, a collision usually damps $i$.
While $\langle i \rangle_M$ generally increases with time, its final value has a large variance. 
The final $\langle i \rangle_M$ is between $10^{-3}$ rad and $10^{-1}$ rad. 

\subsubsection{Dependence of System Properties}\label{sect:e_dependence}

\begin{figure}[htpb]
	\begin{minipage}{0.32\hsize}
	\includegraphics[width=1.2\linewidth]{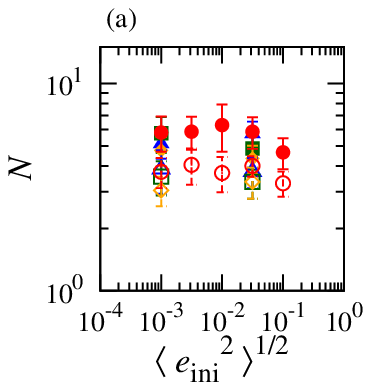}
	\end{minipage}
	\begin{minipage}{0.32\hsize}
	\includegraphics[width=1.2\linewidth]{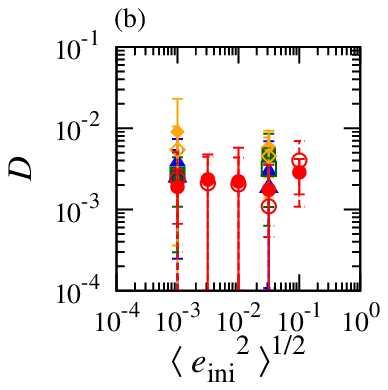}
	\end{minipage}\\
	\begin{minipage}{0.32\hsize}
	\includegraphics[width=1.2\linewidth]{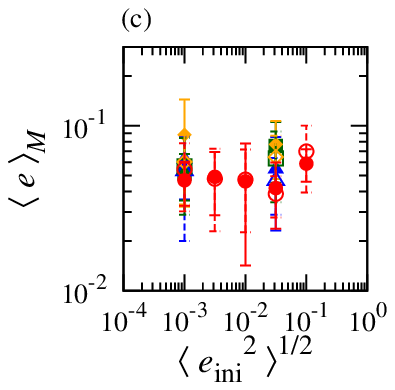}
	\end{minipage}
	\begin{minipage}{0.32\hsize}
	\includegraphics[width=1.2\linewidth]{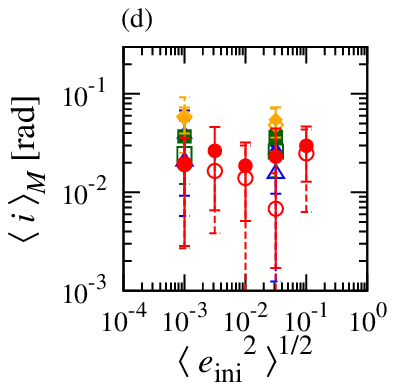}
	\end{minipage}
	\begin{minipage}{0.32\hsize}
	\includegraphics[width=1.2\linewidth]{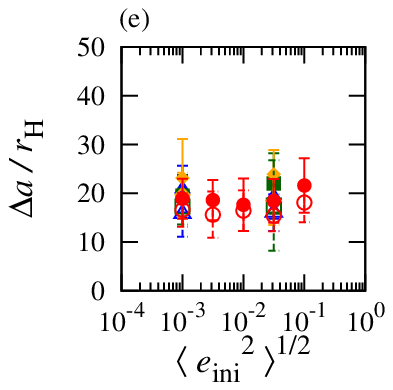}
	\end{minipage}
	\caption{
		Number ($N$; panel a), angular momentum deficit ($D$; b), mass-weighted eccentricity ($\langle e \rangle_M$; c) and inclination
		 ($\langle i \rangle_M$; d), and mean orbital separation normalized by the Hill radius ($\Delta a/r_{\rm H}$;e) of the final planets are plotted
		  as a function of the initial eccentricity dispersion of protoplanets ($\langle e_{\rm ini}^2 \rangle^{1/2}$).
		The error bars indicate the standard deviation.
		The filled symbols are the results of model N16 and the open symbols are those of model N8.
		The initial inclination dispersions are $\langle i_{\rm ini}^2 \rangle^{1/2} = 10^{-3.6}$ rad (red circles),
		 $\langle i_{\rm ini}^2 \rangle^{1/2} = 10^{-2.5}$ rad (blue triangles),
		 $\langle i_{\rm ini}^2 \rangle^{1/2} = 10^{-1.5}$ rad (green squares),
		 and $\langle i_{\rm ini}^2 \rangle^{1/2} = 10^{-1}$ rad (orange diamonds).
	}
	\label{fig:e_ini_fit}
\end{figure}

The mean system properties of planetary systems formed by giant impacts in each model are summarized in Table~\ref{table:results},
 where the angular momentum deficit (AMD) \citep[$D$;][]{Laskar1997} is given by
\begin{eqnarray}
	D = \frac{\sum_j M_j\sqrt{a_j}(1- \sqrt{1-e_j^2}\cos{i_j})}
		{\sum_j M_j\sqrt{a_j}},
	\label{eq:AMD}
\end{eqnarray}
and $\Delta a/r_{\rm H}$ is the orbital separation normalized by the mutual Hill radius. 
In Figure~\ref{fig:e_ini_fit}, the system parameters are plotted as a function of $\langle e_{\rm ini}^2 \rangle^{1/2}$.
We fit these data and find that generally the system parameters barely or only weakly depend on $\langle e_{\rm ini}^2 \rangle^{1/2}$ in both models N16 and N8.

We focus on the results of $\langle i_{\rm ini}^2 \rangle^{1/2} = 10^{-3.6}$ rad (red circles in Figure \ref{fig:e_ini_fit})
 to consider the dependences on $\langle e_{\rm ini}^2 \rangle^{1/2}$.
In the following, when we mention models N16 and N8, models N16e-1i-3.6 and N8e-1i-3.6 are not included. 
In these models, $\langle e_{\rm ini}^2 \rangle^{1/2}\gtrsim e_{\rm c}$, where $e_{\rm c}=0.5\Delta a/a$, and the behaviors of the parameters are different from those
 in the other models.
When $\langle e_{\rm ini}^2 \rangle^{1/2}\gtrsim e_{\rm c}$, final $\langle e\rangle_M$,
 $\langle i\rangle_M$, and $\Delta a/r_{\rm H}$ are larger than those in any other models.
This feature agrees with the results of \cite{Dawson+2016}.

The number of final planets in a system is plotted in panel (a) of Figure \ref{fig:e_ini_fit}.
Although there are some fluctuations, we obtain $N=6$ in model N16, and $N=4$ in model N8, when $\langle e_{\rm ini}^2 \rangle^{1/2}< e_{\rm c}$.
While the final number of planets in model N16 is larger than that in Model N8, the protoplanets in model N16 need more collisions
 than those in model N8. 
This is because less collisions are needed in model N8 to increase orbital separations and become orbitally stable.
These values are not changed as $\langle e_{\rm ini}^2 \rangle^{1/2}$ increases. 
The numbers of final planets in models N16e-1i-3.6 and N8e-1i-3.6 are smaller, $N=4.7$ and 3.3, respectively.

The resultant mass-weighted eccentricities and inclinations are shown in panels (c) and (d) of Figure \ref{fig:e_ini_fit}.
The mass-weighted eccentricities are similar values in all models, $10^{-2}\lesssim \langle e\rangle_M \lesssim10^{-1}$.
This means that $\langle e\rangle_M$ is well relaxed by several collisions. 

The mass-weighted inclinations in model N16 are not changed systematically with $\langle e_{\rm ini}^2 \rangle^{1/2}$, i.e., they are well relaxed.
Their values are $\simeq2\times10^{-2}$ rad. 
Those in model N8 decrease with increasing $\langle e_{\rm ini}^2 \rangle^{1/2}$. 
In the large $\langle e_{\rm ini}^2\rangle^{1/2}$ models, the collision timescale is shorter. 
Protoplanets collide before their inclinations are pumped up, which leads to small $\langle i\rangle_M$ in model N8.
This feature is not clear in model N16. 
The number of collisions in model N16 is about twice that of model N8.
The protoplanets in model N16 suffer more scattering and collisions, which relax the initial conditions. 

Since the masses of planets are similar and their orbits are packed, the AMD (panel (b) of Figure \ref{fig:e_ini_fit}) depends strongly on $e_j$ and $i_j$. 
Under the approximation that $e_j^2\ll1$ and $i_j^2\ll1$, the AMD can be approximated as 
\begin{eqnarray}
	D \simeq \frac{\sum_j 0.5M_j\sqrt{a_j}(e_j^2+i_j^2 )}{\sum_j M_j\sqrt{a_j}},
	\label{eq:AMD_2}
\end{eqnarray}
which can be estimated by $\langle e\rangle_M$ and $\langle i\rangle_M$.

The orbital separations (panel (e) of Figure \ref{fig:e_ini_fit}) in model N16 are about $18r_{\rm H}$, and those in model N8
 are about $16r_{\rm H}$. 
This difference can be interpreted by the orbital crossing time. 
When $N$ is small ($N\lesssim 10$, typically), the orbital crossing time decreases with increasing $N$
 \citep{Chambers+1996, Funk+2010, Matsumoto+2012}, and the orbital crossing time sensitively depends on $e/h$
 \citep{Yoshinaga+1999, Zhou+2007, Pu&Wu2015}. 
The eccentricities of planets are independent of the initial $N$.
The most part of the final $e$ is obtained by the first several collisions (\S \ref{sect:e_init}, Figure \ref{fig:Time_ev_N+massweighted_e_i_runA_e}).
Then the crossing time increases as $\Delta a/r_{\rm H}$ increases by collisions. 
Since $N$ of planets in model N16 is larger than that in model N8, masses of planets in model N16 are smaller, and $h$ is smaller than those in model N8.
The wider separations are needed for a system to be stable in model N16 due to larger $e/h$.

These results would be barely affected by the assumption of perfect accretion. 
Except for models N16e-1i-3.6 and N8e-1i-3.6, where protoplanets initially have $e \sim e_{\rm esc}$, the average collision velocities are
 $1.2v_{\rm esc}$ in model N16 and $1.1v_{\rm esc}$ in model N8, respectively.
These collision velocities are results of collisions between neighboring protoplanets soon after orbital crossing.
Under these collision velocities, over 70\% collisions are merging ones \citep{Kokubo&Genda2010, Genda+2012}.

\subsection{Effect of Initial Inclinations}\label{sect:i}

\begin{figure}[htpb]
	\plotone{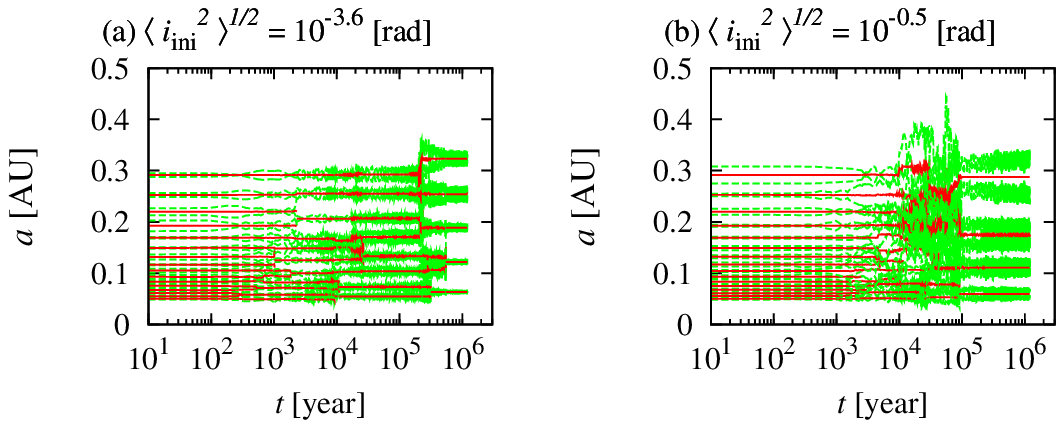}
	\caption{Same as Figure \ref{fig:ap_runA_e}, but for models N16e-1.5i-3.6 (a) and N16e-1.5i-0.5 (b).
	}
	\label{fig:ap_runA_i}
\end{figure}

\begin{figure}[htpb]
	\plotone{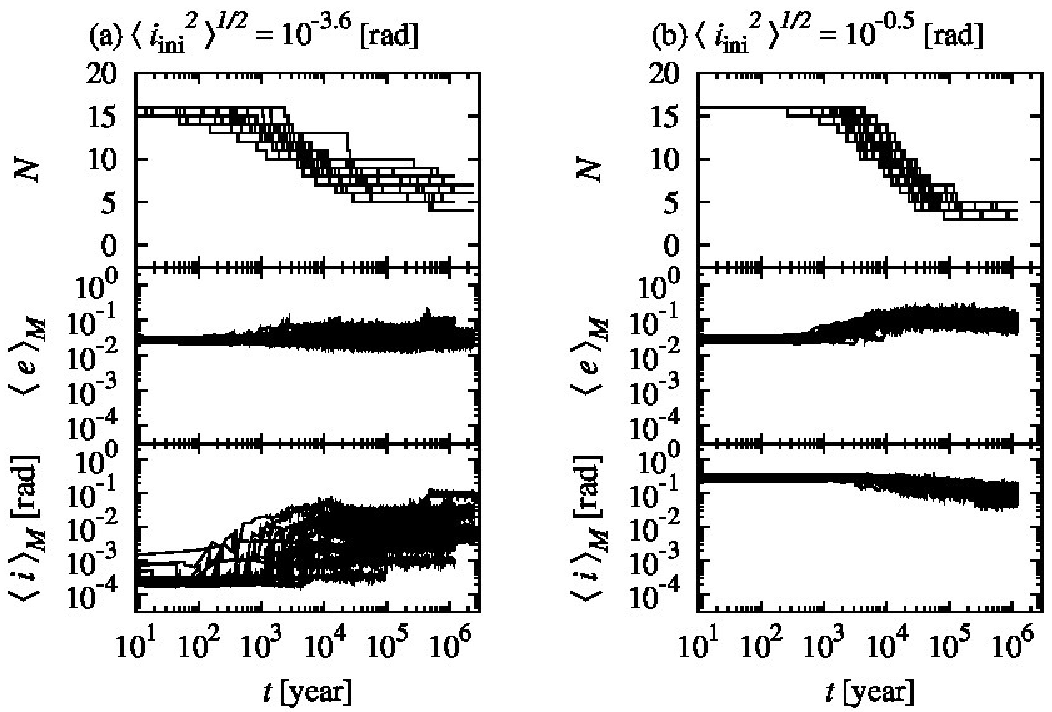}
	\caption{
		Same as Figure \ref{fig:Time_ev_N+massweighted_e_i_runA_e}, but for models N16e-1.5i-3.6 (a) and N16e-1.5i-0.5 (b).
	}
	\label{fig:Time_ev_N+massweighted_e_i_runA_i}
\end{figure}

Next we examine the dependence of the orbital evolution and the final structure of planetary systems on the initial inclinations of protoplanets.
Panels (a) of Figure \ref{fig:ap_runA_i} and \ref{fig:Time_ev_N+massweighted_e_i_runA_i}  show an example of the orbital evolution
 and the time evolution of $N$, $\langle e \rangle_M$ and $\langle i \rangle_M$ of model N16e-1.5i-3.6, where protoplanets have small initial
 dispersions of inclinations.
In this run, $\langle e_{\rm ini}^2 \rangle^{1/2} = 10^{-1.5}$, which corresponds to a few times the reduced Hill radius. 
Since $\langle e_{\rm ini}^2 \rangle^{1/2}$ of protoplanets are not very large, the tournament-like collisional evolution is clearly seen.
This means that protoplanets collide with adjacent protoplanets and collisions occur soon after their orbits are crossed.
The average times of the first and final collisions are $(5.0\pm 7.0)\times10^2$ year and $(2.3\pm 2.8)\times10^5$ year, respectively. 
The time evolution of $N$ in model N16e-1.5i-3.6 is similar to that in model N16e-1i-3.6, although the timescale of $N$ evolution in model N16e-1.5i-3.6
 is longer than that in model N16e-1i-3.6.

\label{sect:i_large}

Panel (b) of Figure \ref{fig:ap_runA_i} shows the orbital evolution of a run of model N16e-1.5i-0.5, where the inclinations are larger than $i_{\rm esc}$.
Protoplanets tend to collide with adjacent protoplanets in the same way as the small inclination model. 
Compared with model N16e-1.5i-3.6, the first collision needs more time to occur and some protoplanets do not collide soon
 after orbital crossing in model N16e-1.5i-0.5. 

Figure \ref{fig:Time_ev_N+massweighted_e_i_runA_i} also shows the time evolution of $N$, $\langle e \rangle_M$,
 and $\langle i \rangle_M$ in all the runs of model N16e-1.5i-0.5.
The first and final collisions occur at $(2.1\pm 0.98)\times10^3$ year and $(2.7\pm 2.5)\times10^5$ year, respectively.
The first collisions in model N16e-1.5i-3.6 occur earlier than those in model N16e-1.5i-0.5.
We find that while the timing of the first collision of small $\langle e_{\rm ini}^2 \rangle^{1/2}$ model ($\langle e_{\rm ini}^2 \rangle^{1/2} = 10^{-3}$)
 decreases with increasing $\langle i_{\rm ini}^2\rangle^{1/2}$, that of large $\langle e_{\rm ini}^2 \rangle^{1/2}$ model
 ($\langle e_{\rm ini}^2 \rangle^{1/2} = 10^{-1.5}$) increases with $\langle i_{\rm ini}^2 \rangle^{1/2}$.
For $\langle i_{\rm ini}^2 \rangle^{1/2}=10^{-1}$ rad, the time of the first collision is comparable between the large and small 
 $\langle e_{\rm ini}^2 \rangle^{1/2}$ models.
These features are caused by the crossing time and relative velocity of protoplanets.
We define the collision timescale as the time between collisions.
When the crossing timescale is a major part of the collision timescale, the first collision time decreases with increasing $\langle i_{\rm ini}^2 \rangle^{1/2}$
 since the crossing time becomes shorter \citep{Pu&Wu2015}.
On the other hand, when the crossing time is a minor fraction in the collision timescale, the first collision time increases with $\langle i_{\rm ini}^2 \rangle^{1/2}$
 since the relative velocities become larger and the collision time becomes longer \citep{Dawson+2016}.
As a result of the relaxation of the initial inclinations through giant impacts, the time of the final collisions tends to be comparable between
 models N16e-1.5i-3.6 and N16e-1.5i-0.5. 
Thus, the accretion timescale, which is estimated as the time difference between the first and final collisions, is not changed by $\langle i_{\rm ini}^2 \rangle^{1/2}$. 

The final $\langle e \rangle_M$ and $\langle i \rangle_M$ are $\sim 10^{-1}$ and $\sim 10^{-1}$ rad in model N16e-1.5i-0.5.
These values are larger than those in model N16e-1.5i-3.6.
For $\langle i_{\rm ini}^2 \rangle^{1/2}\gtrsim i_{\rm esc}$, $\langle i \rangle_M$ decreases gradually by collisional damping
 (\S \ref{sect:col_damp_i}), and finally becomes $\langle i \rangle_M\lesssim i_{\rm esc}$.
The collisional damping of inclinations strongly depends on $\langle i_{\rm ini}^2 \rangle^{1/2}$.
For large $\langle i_{\rm ini}^2 \rangle^{1/2}$, the collisional damping becomes less effective.
The detailed mechanism is discussed in the next section.
This difference of $i$ also changes the other system parameters.
For larger $i$, the collision time is longer and $e$ just before collisions tends to be larger. 
This causes the eccentricity-damping by a collision less effective \citep{Matsumoto+2015}, and the final $\langle e \rangle_M$
 in model N16e-1.5i-0.5 is larger than that in model N16e-1.5i-3.6. 
The final $N$ in model N16e-1.5i-0.5 is smaller than that in model N16e-1.5i-3.6, since $e$ in model N16e-1.5i-3.6 is larger. 

\subsubsection{Orbital Evolution}\label{sect:col_damp_i}

\begin{figure}[htpb]
	\plotone{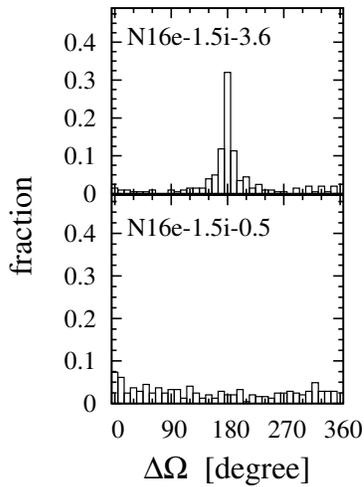}
	\caption{
		Distribution of the difference of the longitude of ascending node ($\Delta \Omega$) of colliding bodies just before the collision
		 in models N16e-1.5i-3.6 (top) and N16e-1.5i-0.5 (bottom).
		The total numbers of collisions are 203 in model N16e-1.5i-3.6 and 245 in model N16e-1.5i-0.5.
		The mean and variance of $\Delta \Omega$ are $184^{\circ}\pm 62.1^{\circ}$ in model N16e-1.5i-3.6 and $166^{\circ}\pm 117^{\circ}$ in model N16e-1.5i-0.5.
	}
	\label{fig:i_ini_vs_dOmega_col}
\end{figure}

When a collision takes place, the inclination of the merged body becomes smaller than the osculating inclinations of the colliding bodies
 \citep{Matsumoto+2015}. 
Since the velocity component normal to the invariable plane ($v_z$) is proportional to $\cos{\Omega}$, where $\Omega$ is
 the ascending node of a body, when the difference between the ascending nodes of colliding bodies $\Delta \Omega$ is
 around $180^{\circ}$ just before a collision, the orbital plane of the merged body becomes closer to the invariable plane.
Figure \ref{fig:i_ini_vs_dOmega_col} shows the $\Delta \Omega$ distribution just before collisions in models N16e-1.5i-3.6 and N16e-1.5i-0.5. 
In model N16e-1.5i-3.6, where $\langle i_{\rm ini}^2 \rangle^{1/2}$ is small, $\Delta \Omega$ is concentrated around $180^{\circ}$,
 while in model N16e-1.5i-0.5, where $\langle i_{\rm ini}^2\rangle^{1/2}$ is large, $\Delta \Omega$ is almost uniform. 

The $\Delta \Omega$ distribution depends on $\langle i_{\rm ini}^2\rangle^{1/2}$. 
This can be explained by the effect of gravitational focusing.
When $\langle i_{\rm ini}^2 \rangle^{1/2}$ is small, $v_z$ of a protoplanet is small and gravitational focusing is effective. 
In this case, due to the mutual gravity between colliding protoplanets, $v_z$ is accelerated. 
When this velocity change is larger than $\langle i_{\rm ini}^2 \rangle^{1/2} v_{\rm K}$, the ascending nodes of the protoplanets
 are changed so that $\Delta \Omega$ becomes about $180^{\circ}$. 
The collisional damping becomes very effective, and most of $i$ gained by gravitational focusing is damped by the collision.
For large $\langle i_{\rm ini}^2 \rangle^{1/2}$, this velocity change is smaller than $\langle i_{\rm ini}^2 \rangle^{1/2} v_{\rm K}$. 
The ascending nodes of the protoplanets are barely changed and $\Delta \Omega$ is not concentrated around $180^{\circ}$. 
Even in this case, $i$ is damped by the collision, though damping is less effective. 

\begin{figure}[htpb]
	\plotone{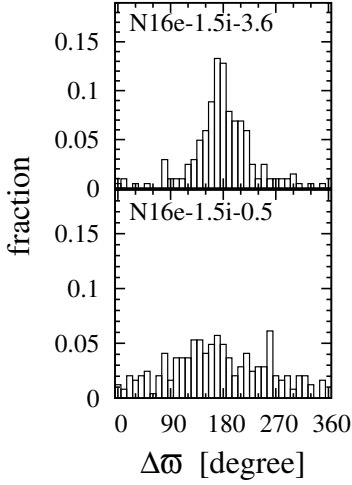}
	\caption{
	Same as Figure \ref{fig:sort_dvarpi_e}, but for models N16e-1.5i-3.6 (top) and N16e-1.5i-0.5 (bottom).
	The mean and variance of $\Delta \varpi$ are $180^{\circ}\pm 52.9^{\circ}$ in model N16e-1.5i-3.6 and $172^{\circ}\pm 83.3^{\circ}$ in model N16e-1.5i-0.5.
	}	
	\label{fig:sort_dvarpi_i}
\end{figure}

Gravitational focusing also affects collisional damping of $e$.
Due to the change of the osculating orbits in a collision, $\Delta \varpi$ is changed just before a collision. 
For $\Delta \varpi \simeq 180^{\circ}$, since the relative velocity is larger, gravitational focusing is less effective. 
Although gravitational focusing is effective for $\Delta \varpi \simeq 0^{\circ}$, it is hard for protoplanets to collide in this configuration. 
Thus gravitational focusing is less important for $\Delta \varpi$ than for $\Delta \Omega$. 
Figure \ref{fig:sort_dvarpi_i} shows the $\Delta \varpi$ distributions just before collisions in models N16e-1.5i-3.6 and N16e-1.5i-0.5. 
We find that $\Delta \varpi$ is more concentrated around $180^{\circ}$ in model N16e-1.5i-3.6.
For larger $\langle i_{\rm ini}^2\rangle^{1/2}$, the relative velocity between protoplanets is largeer and gravitational focusing is less effective,
 which leads to less concentration of $\Delta \varpi$ around $180^{\circ}$.

\subsubsection{Dependence of System Properties}

The mean system properties of planetary systems in each model are summarized in Table~\ref{table:results}. 
Figure \ref{fig:i_ini_fit} shows the system parameter dependence on $\langle i_{\rm ini}^2\rangle^{1/2}$ in the same manner
 as Figure \ref{fig:e_ini_fit}, where the data are fitted by 
\begin{eqnarray}
	\log{y} = C_{i1} \log{\langle i_{\rm ini}^2 \rangle^{1/2}} +C_{i2},
	\label{eq:fit_i}
\end{eqnarray}
using the least-squares fitting.
The best-fit values of $C_{i1}$ and $C_{i2}$ are summarized in Table \ref{table:best-fit}. 

In the following, we consider both $\langle e_{\rm ini}^2 \rangle^{1/2}=10^{-1.5}$ models (red symbols)
 and $\langle e_{\rm ini}^2 \rangle^{1/2}=10^{-3}$ models (blue symbols).
This is because the effect of $\langle e_{\rm ini}^2 \rangle^{1/2}$ is well relaxed and the final values of parameters are not affected
 by $\langle e_{\rm ini}^2 \rangle^{1/2}$.
The final values of parameters depend on $\langle i_{\rm ini}^2 \rangle^{1/2}$ (\S \ref{sect:i_large}).
We describe the dependence on $\langle i_{\rm ini}^2 \rangle^{1/2}$,
 and explain whether the number of collisions changes our results, comparing models N16, N8, and N32, below.

The number of final planets in a system (panel a) decreases with increasing $\langle i_{\rm ini}^2 \rangle^{1/2}$ (\S \ref{sect:i_large}).
While we obtain $\langle N\rangle=5.9$ in model N16e-1.5i-3.6, where $\langle i_{\rm ini}^2 \rangle^{1/2}=10^{-3.6}$, 
 $\langle N\rangle=3.8$ in model N16e-1.5i-0.5, where $\langle i_{\rm ini}^2 \rangle^{1/2}=10^{-0.5}$.
Comparing models N16, N8, and N32, there are more planets in models with the larger number of initial protoplanets.
The numbers of collisions are about 10 in model N16, 4 in model N8, and 23 in model N32.
The coefficients $C_{i1}$ in models N16 and N8 for $N$ are similar values, $C_{i1}\simeq0.04$.
This means that the final number of planets depends on $\langle i_{\rm ini}^2 \rangle^{1/2}$, regardless of the initial number.
As we explained in \S \ref{sect:i_large}, this is because eccentricities and inclinations depend on $\langle i_{\rm ini}^2 \rangle^{1/2}$.

The resultant mass-weighted eccentricities and inclinations increase with $\langle i_{\rm ini}^2 \rangle^{1/2}$.
In the case of $\langle e \rangle_M$, the final values in models N16, N8, and N32 are close.
These values are between 0.04 and 0.11.
The coefficients are approximately the same, $C_{i1}\simeq0.07$ and $C_{i2}\simeq-1$ in models N16 and N8.
Except for models N8e-1.5i-3.6 and N8e-1.5i-2.5, where final $\langle i \rangle_M$ are smaller than that in the other models
 (\S \ref{sect:e_dependence}), the final values in models N16, N8, and N32 are close. 
The final $\langle i \rangle_M$ are between 0.02 and 0.12, which is a wider range than those of $\langle e \rangle_M$.
This means that the final $\langle e \rangle_M$ and $\langle i \rangle_M$ depend on $\langle i_{\rm ini}^2 \rangle^{1/2}$,
 regardless of the number of collisions.
The coefficients $C_{i1}$ are 0.17 in model N16 and 0.22 in model N8.
Since $\langle e \rangle_M$ and $\langle i \rangle_M$ are relaxed through orbital evolution, their coefficients $C_{i1}$ are smaller than 1.
Since $\langle e \rangle_M$ and $\langle i \rangle_M$ depend on $\langle i_{\rm ini}^2 \rangle^{1/2}$, the final $D$ depends on $\langle i_{\rm ini}^2 \rangle^{1/2}$.

The final orbital separations increase as $\langle i_{\rm ini}^2 \rangle^{1/2}$ increases.
We also find that the final $\Delta a/r_{\rm H}$ increases with the initial $N$ because of the orbital instability (\S \ref{sect:e_dependence}).
The final $\Delta a/r_{\rm H}$ in models N16 and N32 increases weakly with $\langle i_{\rm ini}^2 \rangle^{1/2}$.
The final $\Delta a/r_{\rm H}$ in model N8 increases largely at $\langle i_{\rm ini}^2 \rangle^{1/2}=10^{-0.5}$ rad (model N8e-1.5i-0.5).
This large change is due to small final $N$.
As $N$ decreases, the orbital separations become larger and larger.
The final number of planets in model N8e-1.5i-0.5 is $2.6\pm0.50$, which means about half the systems have only two planets,
 while the final $N$ in the other models is more than 3.
In the close-in orbits, the ranges of the semimajor axes are hardly changed by collisions, since the gravity of the central star is strong.
In such a situation, $\langle \Delta a \rangle \sim (a_N-a_1)/N$, where $a_N$ is the semimajor axis of the outermost planet,
 and this means that when $N$ is small, $\Delta a$ is largely changed with changing $N$.
This is why the final $\Delta a/r_{\rm H}$ in model N8e-1.5i-0.5 increases largely.

\section{SUMMARY AND DISCUSSION}\label{sect:summary}

We have investigated the effects of the initial eccentricities and inclinations of protoplanets on the formation of close-in planets
 by giant impacts using $N$-body simulations.
The eccentricity and inclination dispersions, $\langle e_{\rm ini}^2 \rangle^{1/2}$ and $\langle i_{\rm ini}^2 \rangle^{1/2}$,
 of protoplanets were independently changed.
We also changed the number of initial protoplanets while fixing the total mass $M_{\rm tot}$ and initial mean semimajor axis $\langle a \rangle$.
The number of collisions on the giant impact stage changes with the number of initial protoplanets.
At this stage scattering and collisions work as the relaxation process of the eccentricity and inclination. 
In the following, we mainly describe the results of the system, where most protoplanets experience at least one collision,
 and their eccentricities and inclinations are relaxed by close scattering and collisions.
We found that in the range of parameters we adopted, the initial eccentricities of protoplanets do not affect the properties of
 the final planetary systems, while the initial inclinations do.

The evolution of protoplanets' eccentricities under the condition that protoplanets are not initially on crossing orbits is summarized as follows:
When the radial excursion of protoplanets $a\langle e_{\rm ini}^2 \rangle^{1/2}$ is smaller than the initial orbital separations,
 the eccentricities are pumped up by mutual scattering.
The osculating eccentricities of colliding protoplanets become $\sim e_{\rm esc}$, and are eventually damped by the collision.
Through the repetition of these processes, the final eccentricities are determined.
Since the radial excursions of the merged protoplanets becomes smaller than their orbital separations, the eccentricity evolution becomes
 similar to the small $a\langle e_{\rm ini}^2 \rangle^{1/2}$ case.
In other words, the difference in the initial eccentricities is relaxed by scattering and collisions.
As a result, the final number of planets $N$, angular momentum deficit $D$, mass-weighted eccentricity $\langle e \rangle_M$ and
 inclination $\langle i \rangle_M$, and orbital separation normalized by the Hill radius $\Delta a/r_{\rm H}$ do not depend
 on $\langle e_{\rm ini}^2 \rangle^{1/2}$.
When the initial orbital excursions of protoplanets are larger than the initial separations,
 the final parameters of planets are different from those in the small $a\langle e_{\rm ini}^2 \rangle^{1/2}$ models.
These features are consistent with \cite{Dawson+2016}.

On the other hand, $\langle i_{\rm ini}^2 \rangle^{1/2}$ is not always relaxed at this stage.
When $\langle i_{\rm ini}^2 \rangle^{1/2} \lesssim h$, the inclination is not always pumped up to $\sim i_{\rm esc}$ by close scattering, 
 since gravitational scattering takes place in the disk plane.
When $\langle i_{\rm ini}^2 \rangle^{1/2} \lesssim i_{\rm esc}$,
 since colliding protoplanets have smaller velocity components normal to the invariant plane than the escape velocity,
 gravitational focusing is effective and their ascending and descending nodes approaches the collision point just before a collision. 
Since the collision with this orbital configuration effectively damps the inclination, the inclination of the merged protoplanet becomes small.
Thus the final planets tend to have small inclinations.
However, when $\langle i_{\rm ini}^2 \rangle^{1/2}$ is larger than $\sim i_{\rm esc}$, collisions occur not only around the ascending
 and descending nodes of colliding protoplanets.
In this case collisional damping of the inclination is not always effective and the inclinations tend to be kept large.
The large initial inclinations also affect $N$, $D$, $\langle e \rangle_M$, $\langle i \rangle_M$, and $\Delta a/r_{\rm H}$ of the final planets.
For larger $\langle i_{\rm ini}^2 \rangle^{1/2}$, collisions need more time to occur.
Thus the eccentricities of protoplanets are well pumped up.
Since large eccentricities cause more collisions, $N$ becomes smaller and $\Delta a/r_{\rm H}$ becomes larger.
The dependence of $\Delta a/r_{\rm H}$ on $\langle i_{\rm ini}^2 \rangle^{1/2}$ agrees with the estimation by \cite{Dawson+2016}.
They estimated $\Delta a/r_{\rm H}$ by the equilibrium between gravitational scatterings and collisions.  

\begin{figure}[htpb]
	\begin{minipage}{0.32\hsize}
	\includegraphics[width=1.2\linewidth]{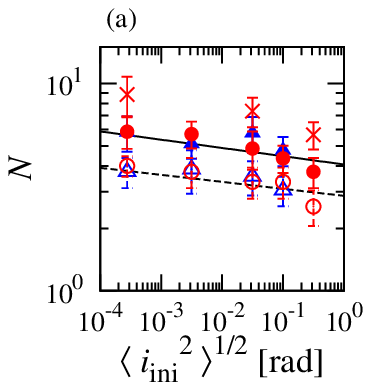}
	\end{minipage}
	\begin{minipage}{0.32\hsize}
	\includegraphics[width=1.2\linewidth]{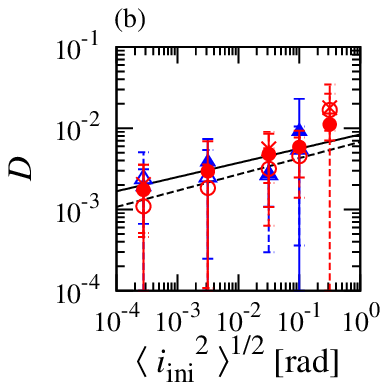}
	\end{minipage}\\
	\begin{minipage}{0.32\hsize}
	\includegraphics[width=1.2\linewidth]{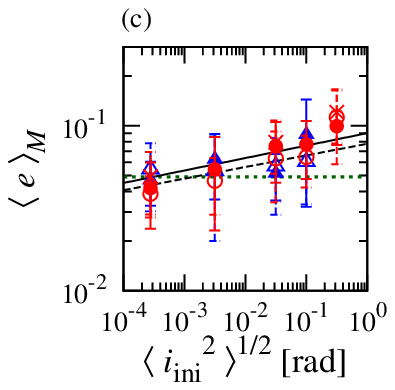}
	\end{minipage}
	\begin{minipage}{0.32\hsize}
	\includegraphics[width=1.2\linewidth]{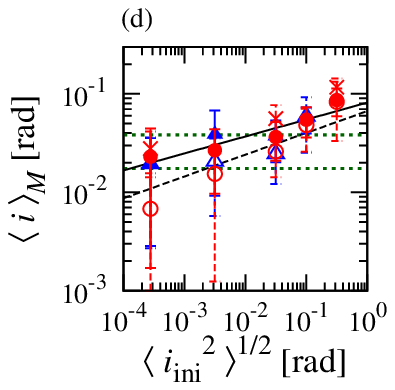}
	\end{minipage}
	\begin{minipage}{0.32\hsize}
	\includegraphics[width=1.2\linewidth]{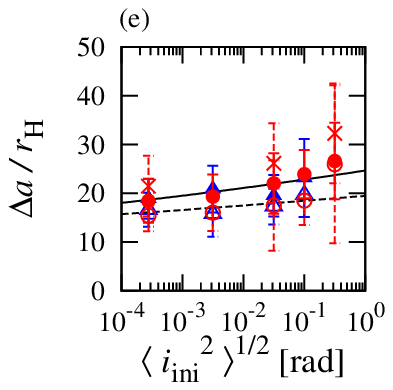}
	\end{minipage}
	\caption{
		Same as Figure \ref{fig:e_ini_fit}, but as a function of the initial inclination dispersion of protoplanets ($\langle i_{\rm ini}^2 \rangle^{1/2}$).
		The cross symbols are the results of model N32.
		The initial eccentricity dispersions are $\langle e_{\rm ini}^2 \rangle^{1/2} = 10^{-1.5}$ rad (red circles),
		 and $\langle e_{\rm ini}^2 \rangle^{1/2} = 10^{-3}$ rad (blue triangles).
		The dotted lines are the best-fit eccentricity ($e=0.049$) and the mode of the mutual inclination (0.017 rad $<i<$ 0.039 rad) of
		 the $Kepler$ planets \citep{Fabrycky+14, Hadden&Lithwick2014, VEylen&Albrecht2015}.
		The solid and dashed lines are the fits by the least-squares-fit method for models N16 and N8 using Equation (\ref{eq:fit_i}).
		The coefficients are summarized in Table \ref{table:best-fit}.
		}
	\label{fig:i_ini_fit}
\end{figure}

We apply our results to the {\it Kepler} planets assuming that they are formed through giant impacts of protoplanets in a gas-free disk.
Substituting the disk surface density given by \cite{Chiang&Laughlin2013} or \cite{Schlichting2014} into the equation of the isolation mass
 \citep{Kokubo&Ida02}, we obtain the mass of protoplanets needed to form close-in super-Earths $\gtrsim M_{\oplus}$.
The initial eccentricities and inclinations of protoplanets depend on the surface density of planetesimals \citep{Kokubo&Ida02}.
\cite{Kokubo&Ida02} showed that protoplanets with $1-5$ Earth mass have eccentricities $\simeq 0.05$ in the $\Sigma_1=100$ model around 1 au,
 which indicates that they have $\simeq 0.025$ rad inclinations.
However we do not know the inclination of protoplanets on close-in orbits ($a \lesssim 0.3$ au).
We find that the eccentricities and inclinations of the {\it Kepler} planets are consistent with our results, if protoplanets initially have
 $\sim10^{-3}-10^{-2}$ rad inclinations (Figure \ref{fig:i_ini_fit}).

In the present paper, we assumed the perfect accretion, which is not always satisfied on the giant impact stage.
Since collisions play an important role in the relaxation process, it is important to understand the outcome of collisions.
The collision velocity increases as $\langle i_{\rm ini}^2 \rangle^{1/2}$ increases.
From the merging criteria given by \cite{Kokubo&Genda2010} and \cite{Genda+2012}, most collisions between protoplanets
 in the $\langle i_{\rm ini}^2\rangle^{1/2}< 10^{-1}$ rad models are not hit-and-run but merging.
Thus the perfect accretion approximation is justified except for the unrealistically large $\langle i_{\rm ini}^2\rangle^{1/2}$ models.

\acknowledgments
We thank the anonymous referee for helpful comments.
Numerical computations were carried out on the PC cluster at the Center for Computational Astrophysics, National Astronomical Observatory of Japan.

\clearpage

\clearpage

%
\begin{deluxetable}{lcccccccccc}
	\tablecaption{Initial Conditions of Protoplanets\label{table:initial}}
	\tablenum{1}
	\renewcommand{\arraystretch}{1}
\tablehead{ 
Model\ &\ $N$\ &\ $\Sigma_1\ $ \ &\ $a_{\rm 1}$ \ & &\ $\langle e_{\rm ini}^2 \rangle^{1/2}$\ &\ $\langle i_{\rm ini}^2 \rangle^{1/2}$\ & &\  $M_{\rm tot}$\ &\ $\langle a \rangle$ \ \ \\
& & $({\rm g\ cm^{-2}})$ &(au) & & & (rad) & & ($M_{\oplus}$) & (au)
}
\startdata
	N16e-1i-3.6	& 16 & 100 & 0.05 & & $10^{-1}$ & $10^{-3.6}$ & & 17.3 & 0.13 \\
	N16e-1.5i-3.6	& 16 & 100 & 0.05 & & $10^{-1.5}$ & $10^{-3.6}$ & & 17.3 & 0.13 \\
	N16e-2i-3.6	& 16 & 100 & 0.05 & & $10^{-2}$ & $10^{-3.6}$ & & 17.3 & 0.13 \\
	N16e-2.5i-3.6	& 16 & 100 & 0.05 & & $10^{-2.5}$ & $10^{-3.6}$ & & 17.3 & 0.13 \\
	N16e-3i-3.6	& 16 & 100 & 0.05 & & $10^{-3}$ & $10^{-3.6}$ & & 17.3 & 0.13 \\
	N16e-1.5i-0.5	& 16 & 100 & 0.05 & & $10^{-1.5}$ & $10^{-0.5}$ & & 17.3 & 0.13 \\
	N16e-1.5i-1	& 16 & 100 & 0.05 & & $10^{-1.5}$ & $10^{-1}$ & & 17.3 & 0.13 \\
	N16e-1.5i-1.5	& 16 & 100 & 0.05 & & $10^{-1.5}$ & $10^{-1.5}$ & & 17.3 & 0.13 \\
	N16e-1.5i-2.5	& 16 & 100 & 0.05 & & $10^{-1.5}$ & $10^{-2.5}$ & & 17.3 & 0.13 \\
	N16e-3i-1	 	& 16 & 100 & 0.05 && $10^{-3}$ & $10^{-1}$ & & 17.3 & 0.13 \\
	N16e-3i-1.5	& 16 & 100 & 0.05 && $10^{-3}$ & $10^{-1.5}$ & & 17.3 & 0.13 \\
	N16e-3i-2.5	& 16 & 100 & 0.05 && $10^{-3}$ & $10^{-2.5}$ & & 17.3 & 0.13\\   \hline
	N8e-1i-3.6	& 8 & 159 & 0.075 & & $10^{-1}$ & $10^{-3.6}$ & & 17.4 & 0.13 \\
	N8e-1.5i-3.6	& 8 & 159 & 0.075 & & $10^{-1.5}$ & $10^{-3.6}$ & & 17.4 & 0.13 \\
	N8e-2i-3.6	& 8 & 159 & 0.075 & & $10^{-2}$ & $10^{-3.6}$ & & 17.4 & 0.13 \\
	N8e-2.5i-3.5	& 8 & 159 & 0.075 & & $10^{-2.5}$ & $10^{-3.6}$ & & 17.4 & 0.13 \\
	N8e-3i-3.6	& 8 & 159 & 0.075 & & $10^{-3}$ & $10^{-3.6}$ & & 17.4 & 0.13 \\
	N8e-1.5i-0.5	& 8 & 159 & 0.075 & & $10^{-1.5}$ & $10^{-0.5}$ & & 17.4 & 0.13 \\
	N8e-1.5i-1	& 8 & 159 & 0.075 & & $10^{-1.5}$ & $10^{-1}$ & & 17.4 & 0.13 \\
	N8e-1.5i-1.5	& 8 & 159 & 0.075 & & $10^{-1.5}$ & $10^{-1.5}$ & & 17.4 & 0.13 \\
	N8e-1.5i-2.5	& 8 & 159 & 0.075 & & $10^{-1.5}$ & $10^{-2.5}$ & & 17.4 & 0.13 \\
	N8e-3i-1	 	& 8 & 159 & 0.075 & & $10^{-3}$ & $10^{-1}$ & & 17.4 & 0.13 \\
	N8e-3i-1.5	& 8 & 159 & 0.075 & & $10^{-3}$ & $10^{-1.5}$ & & 17.4 & 0.13 \\
	N8e-3i-2.5	& 8 & 159 & 0.075 & & $10^{-3}$ & $10^{-2.5}$ & & 17.4 & 0.13 \\  \hline
	N32e-1.5i-3.6	& 32 & 65 & 0.026 & & $10^{-1.5}$ & $10^{-3.6}$ & & 17.4 & 0.13 \\
	N32e-1.5i-1.5	& 32 & 65 & 0.026 & & $10^{-1.5}$ & $10^{-1.5}$ & & 17.4 & 0.13 \\
	N32e-1.5i-0.5	& 32 & 65 & 0.026 & & $10^{-1.5}$ & $10^{-0.5}$ & & 17.4 & 0.13 
\enddata
	\tablecomments{
		Number of protoplanets ($N$), surface density of protoplanets at 1 au ($\Sigma_1$), semimajor axis of the innermost protoplanet ($a_1$),
		 initial eccentricity dispersion ($\langle e_{\rm ini}^2 \rangle^{1/2}$) and initial inclination dispersion ($\langle i_{\rm ini}^2 \rangle^{1/2}$).
		The total mass of a system ($M_{\rm tot}$), and the mean semimajor axis ($\langle a \rangle$) are given from above parameters.
	}
\end{deluxetable}

\clearpage

\begin{deluxetable}{lcccccccccc}
	\tablecaption{Final Planetary Systems\label{table:results}}
	\tablenum{2}
	\renewcommand{\arraystretch}{1.1}
\tablehead{ 
Model\ &\ $\langle N\rangle$\ &\ $\langle D\rangle$ \ &\ $\langle e\rangle_M$\ &\ $\langle i\rangle_M$\ &\ $\langle \Delta a/r_{\rm H}\rangle$\ & \\
&&($10^{-3}$)&($10^{-2}$)&($10^{-2}$ rad)&&
}
\startdata
	N16e-1i-3.6	&  $4.7  \pm 0.79$   &  $2.9 \pm 1.3$   &  $5.9 \pm 1.3 $   &  $3.0 \pm 1.7 $   &  $21.6 \pm 5.6$   \\
	N16e-1.5i-3.6	&  $5.9  \pm 1.01$   &  $1.7 \pm 1.8$   &  $4.2 \pm 1.8 $   &  $2.3 \pm 2.1 $   &  $18.4 \pm 4.5$   \\
	N16e-2i-3.6	&  $6.3  \pm 1.61$   &  $2.2 \pm 3.6$   &  $4.6\pm 3.2 $   &  $1.9 \pm 1.4 $   &  $17.6 \pm 5.4$   \\
	N16e-2.5i-3.6	&  $5.9  \pm 0.96$   &  $2.3 \pm 2.4$   &  $4.9 \pm 2.0 $   &  $2.6 \pm 2.0 $   &  $18.6 \pm 4.1$   \\
	N16e-3i-3.6	&  $5.8  \pm 1.12$   &  $1.9 \pm 1.2$   &  $4.7\pm 1.4 $   &  $1.9 \pm 1.6 $   &  $18.9 \pm 4.1$   \\
	N16e-1.5i-0.5	&  $3.8  \pm 0.62$   &  $11.1\pm 4.1$   &  $9.9 \pm 2.3 $   &  $8.7 \pm 2.7 $   &  $26.6 \pm 7.8$   \\
	N16e-1.5i-1	&  $4.4  \pm 0.65$   &  $5.9 \pm 3.4$   &  $7.7 \pm 3.0 $   &  $5.5 \pm 1.9 $   &  $23.9 \pm 4.9$   \\
	N16e-1.5i-1.5	&  $4.9  \pm 1.01$   &  $4.8 \pm 3.7$   &  $7.5 \pm 3.0 $   &  $3.6 \pm 1.5 $   &  $22.0 \pm 6.2$   \\
	N16e-1.5i-2.5	&  $5.7  \pm 0.84$   &  $3.0 \pm 4.0$   &  $5.4 \pm 3.1 $   &  $2.7 \pm 1.7 $   &  $19.3 \pm 4.6$   \\
	N16e-3i-1		&  $4.8  \pm 0.77$   &  $9.1 \pm 13.9$   &  $8.8 \pm 5.6 $   &  $5.6 \pm 1.7 $   &  $23.1\pm 8.0$ \\
	N16e-3i-1.5	& $ 5.8 \pm 1.13 $ & $2.7 \pm 1.7$ & $ 5.1 \pm 1.6  $ & $3.7 \pm 1.7  $ & $ 19.5 \pm 4.3 $ \\
	N16e-3i-2.5	& $ 5.1 \pm 0.77 $ & $3.8 \pm 3.6$ & $ 6.2 \pm 2.7  $ & $3.9 \pm 2.9  $ & $ 20.8 \pm 4.8 $ \\ \hline
	N8e-1i-3.6	& $ 3.3 \pm 0.46 $ & $4.0 \pm 2.9$ & $7.0 \pm 3.0  $ & $2.5 \pm 1.9  $ & $ 18.1 \pm 4.0 $ \\
	N8e-1.5i-3.6	& $ 4.0 \pm 0.45 $ & $1.1 \pm 0.63$ & $3.9 \pm 1.1  $ & $0.68 \pm 0.88  $ & $ 15.4 \pm 3.2$ \\
	N8e-2i-3.6	& $ 3.7 \pm 0.71 $ & $2.0 \pm 2.3$ & $4.7 \pm 2.5  $ & $1.4 \pm 1.6  $ & $ 16.4 \pm 4.2 $ \\
	N8e-2.5i-3.5	& $ 4.1 \pm 0.80 $ & $2.1 \pm 2.3$ & $4.8 \pm 2.5  $ & $1.7 \pm 1.3  $ & $ 15.6 \pm 4.7$ \\
	N8e-3i-3.6	& $ 3.8 \pm 0.62 $ & $2.4 \pm 2.7$ & $5.4 \pm 2.4  $ & $1.9 \pm 1.7  $ & $ 16.6 \pm 3.5 $ \\
	N8e-1.5i-0.5	& $ 2.6 \pm 0.50 $ & $17 \pm 18$ & $11.2 \pm 5.4  $ & $8.3 \pm 5.0  $ & $ 25.9 \pm 16.2$ \\
	N8e-1.5i-1	& $ 3.4 \pm 0.57 $ & $4.5 \pm 3.1$ & $6.5 \pm 2.2  $ & $4.8 \pm 2.2  $ & $ 18.5 \pm 5.0 $ \\
	N8e-1.5i-1.5	& $ 3.4 \pm 0.57 $ & $3.2 \pm 2.5$ & $6.3 \pm 2.9  $ & $2.6 \pm 1.2  $ & $ 17.5 \pm 9.3 $ \\
	N8e-1.5i-2.5	& $ 3.8 \pm 0.62 $ & $1.8 \pm 1.7$ & $4.6 \pm 1.8  $ & $1.5 \pm 1.4  $ & $ 16.0 \pm 3.8 $ \\
	N8e-3i-1		& $ 3.1 \pm 0.50 $ & $5.4 \pm 5.1$ & $6.0 \pm 2.7  $ & $5.9 \pm 3.4  $ & $ 19.6 \pm 4.5$ \\
	N8e-3i-1.5	& $ 3.6 \pm 0.67 $ & $2.7 \pm 2.4$ & $5.7 \pm 2.8  $ & $2.5 \pm 1.2  $ & $ 17.3 \pm 3.7 $\\
	N8e-3i-2.5	& $ 3.9 \pm 0.91 $ & $2.5 \pm 2.9$ & $5.3 \pm 3.3  $ & $2.1 \pm 1.5  $ & $ 15.7 \pm 4.6$\\ \hline
	N32e-1.5i-3.6	& $ 8.9 \pm 1.9 $  & $2.0 \pm 1.5$   & $4.9 \pm 2.0  $  & $2.8 \pm 1.4  $ & $ 21.4 \pm 6.3 $ \\
	N32e-1.5i-1.5	& $ 7.4 \pm 1.2$   & $5.6 \pm 3.5$   & $7.9 \pm 2.8  $  & $5.6 \pm 2.1  $ & $ 26.2 \pm 8.2$ \\
	N32e-1.5i-0.5	& $ 5.7 \pm 0.85$ & $17.9 \pm 8.6$ & $12.1 \pm 4.3$ & $11.7 \pm 2.7  $ & $ 32.3 \pm 10.2$ \\
 \enddata
	\tablecomments{
		Final number of planets ($N$), angular momentum deficits ($D$), mass-weighted eccentricities ($\langle e \rangle_M$), 
		 mass-weighted inclinations ($\langle i \rangle_M$), and the orbital separations normalized by the Hill radius ($\Delta a/r_{\rm H}$).
		}
\end{deluxetable}

\clearpage

\begin{deluxetable}{cccc}
	\tabletypesize{\small}
	\tablecaption{Best-Fit Coefficients and their dispersions of Equation (\ref{eq:fit_i}) \label{table:best-fit}}
	\tablenum{3}
	\renewcommand{\arraystretch}{1}
\tablehead{ 
Model & & N16 & N8
}
\startdata
 $N$ & $C_{i1}$
 & $(-4.0\pm 1.2)\times10^{-2}$ & $(-3.4\pm 0.98)\times10^{-2}$ \\ 
 & $C_{i2}$
 & $0.61\pm 3.0\times10^{-2}$ & $0.46\pm 2.7\times10^{-2}$ \\  \ \\
 $D$ & $C_{i1}$
 & $0.17\pm 4.3\times10^{-2}$ & $0.20\pm 4.8\times10^{-2}$  \\   
 & $C_{i2}$
 & $-2.1\pm 0.10$ & $-2.2\pm 0.12$ \\   \ \\
 $\langle e \rangle_M$ & $C_{i1}$
 & $(7.5 \pm 1.9)\times10^{-2}$ & $(7.0\pm 2.1)\times10^{-2}$  \\   
 & $C_{i2}$
 & $-1.0\pm 5.0\times10^{-2}$ & $-1.1\pm 5.7\times10^{-2}$ \\ \ \\
 $\langle i \rangle_M$ & $C_{i1}$
 & $0.17\pm 2.7\times10^{-2}$ & $0.22 \pm 5.2\times10^{-2}$  \\   
 & $C_{i2}$
 & $-1.1\pm 4.8\times10^{-2}$ & $-1.2\pm 0.10$ \\ \ \\
 $\Delta a/r_{\rm H}$ & $C_{i1}$
 & $(3.4\pm 0.85)\times10^{-2}$ & $(2.3\pm 0.86)\times10^{-2}$  \\   
 & $C_{i2}$
 & $1.4 \pm 2.2\times10^{-2}$ & $1.3\pm 2.4\times10^{-2}$ \\ 
\enddata
\end{deluxetable}


\begin{thebibliography}{}

\bibitem[Baruteau \& Papaloizou(2013)]{Baruteau&Papaloizou2013} Baruteau, C., \& Papaloizou, J.~C.~B.\ 2013, \apj, 778, 7 

\bibitem[Chambers et al.(1996)]{Chambers+1996} Chambers, J.~E., Wetherill, G.~W., \& Boss, A.~P.\ 1996, \icarus, 119, 261 

\bibitem[Chiang \& Laughlin(2013)]{Chiang&Laughlin2013} Chiang, E., \& Laughlin, G.\ 2013, \mnras, 431, 3444

\bibitem[Dawson et al.(2016)]{Dawson+2016} Dawson, R.~I., Lee, E.~J., \& Chiang, E.\ 2016, \apj, 822, 54 

\bibitem[Fabrycky et al.(2014)]{Fabrycky+14} Fabrycky, D.~C., Lissauer, J.~J., Ragozzine, D., et al.\ 2014, \apj, 790, 146

\bibitem[Funk et al.(2010)]{Funk+2010} Funk, B., Wuchterl, G., Schwarz, R., Pilat-Lohinger, E., \& Eggl, S.\ 2010, \aap, 516, A82 

\bibitem[Genda et al.(2015)]{Genda+2015} Genda, H., Kobayashi, H., \& Kokubo, E.\ 2015, \apj, 810, 136 

\bibitem[Genda et al.(2012)]{Genda+2012} Genda, H., Kokubo, E., \& Ida, S.\ 2012, \apj, 744, 137 

\bibitem[Hadden \& Lithwick(2014)]{Hadden&Lithwick2014} Hadden, S., \& Lithwick, Y.\ 2014, \apj, 787, 80 

\bibitem[Hansen \& Murray(2012)]{Hansen&Murray12} Hansen, B.~M.~S., \& Murray, N.\ 2012, \apj, 751, 158 

\bibitem[Hansen \& Murray(2013)]{Hansen&Murray13} Hansen, B.~M.~S., \& Murray, N.\ 2013, \apj, 775, 53 

\bibitem[Ida(1990)]{Ida1990} Ida, S.\ 1990, \icarus, 88, 129 

\bibitem[Ida \& Lin(2010)]{Ida&Lin2010} Ida, S., \& Lin, D.~N.~C.\ 2010, \apj, 719, 810 

\bibitem[Ida \& Makino(1992)]{Ida&Makino1992} Ida, S., \& Makino, J.\ 1992, \icarus, 96, 107 

\bibitem[Kokubo \& Genda(2010)]{Kokubo&Genda2010} Kokubo, E., \& Genda, H.\ 2010, \apjl, 714, L21

\bibitem[Kokubo \& Ida(2002)]{Kokubo&Ida02} Kokubo, E., \& Ida, S.\ 2002, \apj, 581, 666

\bibitem[Kokubo \& Ida(2007)]{Kokubo&Ida07} Kokubo, E., \& Ida, S.\ 2007, \apj, 671, 2082 

\bibitem[Kokubo \& Makino(2004)]{Kokubo&Makino2004} Kokubo, E., \& Makino, J.\ 2004, \pasj, 56, 861 

\bibitem[Laskar(1997)]{Laskar1997} Laskar, J.\ 1997, \aap, 317, L75 

\bibitem[Lissauer \& Stewart(1993)]{Lissauer&Stewart1993} Lissauer, J.~J., \& Stewart, G.~R.\ 1993, Protostars and Planets III, 1061 

\bibitem[Lissauer et al.(2011)]{Lissauer+2011} Lissauer, J.~J., Ragozzine, D., Fabrycky, D.~C., et al.\ 2011, \apjs, 197, 8 

\bibitem[Makino(1991)]{Makino1991} Makino, J.\ 1991, \pasj, 43, 859 

\bibitem[Matsumoto et al.(2012)]{Matsumoto+2012} Matsumoto, Y., Nagasawa, M., \& Ida, S.\ 2012, \icarus, 221, 624

\bibitem[Matsumoto et al.(2015)]{Matsumoto+2015} Matsumoto, Y., Nagasawa, M., \& Ida, S.\ 2015, \apj, 810, 106 

\bibitem[Mayor et al.(2011)]{Mayor+2011} Mayor, M., Marmier, M., Lovis, C., et al.\ 2011, arXiv:1109.2497

\bibitem[Moriarty \& Ballard(2016)]{Moriarty&Ballard2016} Moriarty, J., \& Ballard, S.\ 2016, \apj, 832, 34 

\bibitem[Mullally et al.(2015)]{Mullalley+2015} Mullally, F., Coughlin, J.~L., Thompson, S.~E., et al.\ 2015, \apjs, 217, 31 

\bibitem[Ogihara et al.(2015a)]{Ogihara+2015} Ogihara, M., Morbidelli, A., \& Guillot, T.\ 2015, \aap, 578, A36 
 
\bibitem[Ogihara et al.(2015b)]{Ogihara+2015b} Ogihara, M., Morbidelli, A., \& Guillot, T.\ 2015, \aap, 584, L1 

\bibitem[Pu \& Wu(2015)]{Pu&Wu2015} Pu, B., \& Wu, Y.\ 2015, \apj, 807, 44 

\bibitem[Schlichting(2014)]{Schlichting2014} Schlichting, H.~E.\ 2014, \apjl, 795, L15 

\bibitem[Stewart \& Leinhardt(2012)]{Stewart&Leinhardt2012} Stewart, S.~T., \& Leinhardt, Z.~M.\ 2012, \apj, 751, 32 

\bibitem[Tremaine(2015)]{Tremaine2015} Tremaine, S.\ 2015, \apj, 807, 157 

\bibitem[Van Eylen \& Albrecht(2015)]{VEylen&Albrecht2015} Van Eylen, V., \& Albrecht, S.\ 2015, \apj, 808, 126 

\bibitem[Yoshinaga et al.(1999)]{Yoshinaga+1999} Yoshinaga, K., Kokubo, E., \& Makino, J.\ 1999, \icarus, 139, 328 

\bibitem[Zhou et al.(2007)]{Zhou+2007} Zhou, J.-L., Lin, D.~N.~C., \& Sun, Y.-S.\ 2007, \apj, 666, 423 

\end{thebibliography}
\end{document}